\def\year{2018}\relax
\documentclass[letterpaper]{article} 
\usepackage{aaai18}  
\usepackage{times}  
\usepackage{helvet}  
\usepackage{courier}  
\usepackage{url}  
\usepackage{graphicx}  
\usepackage{amsmath}
\usepackage{multirow}
\usepackage[linesnumbered,ruled]{algorithm2e}
\begin{document}
%

\title{Collaborative Filtering with Social Exposure: \\A Modular Approach to Social Recommendation}
\author{
Menghan Wang \and Xiaolin Zheng\thanks{Corresponding author} \and Yang Yang\\
College of Computer Science,\\
Zhejiang University, Hangzhou, 310027, China  \\
\{wangmengh, xlzheng, yangya\}@zju.edu.cn\\
\And Kun Zhang\\
Department of Philosophy,\\
Carnegie Mellon University, Pittsburgh, USA \\
kunz1@cmu.edu\\
}
\maketitle
\begin{abstract}
This paper is concerned with how to make efficient use of social information to improve recommendations. Most existing social recommender systems assume people share similar preferences with their social friends. Which, however, may not hold true due to various motivations of making online friends and dynamics of online social networks. Inspired by recent causal process based recommendations that first model user exposures towards items and then use these exposures to guide rating prediction, we utilize social information to capture user exposures rather than user preferences. We assume that people get information of products from their online friends and they do not have to share similar preferences, which is less restrictive and seems closer to reality. Under this new assumption, in this paper, we present a novel recommendation approach (named SERec) to integrate social exposure into collaborative filtering. We propose two methods to implement SERec, namely \emph{social regularization} and \emph{social boosting}, each with different ways to construct social exposures. Experiments on four real-world datasets demonstrate that our methods outperform the state-of-the-art methods on top-N recommendations. Further study compares the robustness and scalability of the two proposed methods.
\end{abstract}

\section{Introduction}
Social recommendation has became a hot research area, and many social recommender systems have been proposed in recent years.
Most existing social recommender systems assumed users who are connected are more likely to have similar preferences. This widely used assumption can be explained by social correlation theories such as social influence \cite{Marsden1993Network} and homophily \cite{Mcpherson2001Birds}. Under this assumption, social information is always utilized to better capture user latent preferences for collaborative filtering (CF), which is a popular technique used by recommender systems.

However, this assumption should be further examined because the increasing popularity of social media makes the situation more complex. The user characteristics change dynamically and why users interact in the social networks become diverse. It becomes easier to make friends today over the Internet and people make online friends for various reasons: e.g., alumni, colleagues, living in the same city, sharing similar interests. Online friends may not share similarities in preference. In fact, social network researchers have noticed similar phenomenon and studied dynamics of social networks such as information diffusion \cite{Hu2015Information}, social contagion \cite{DBLP:conf/aaai/YangJWT16}, and social structural influence \cite{DBLP:conf/aaai/ZhangTZMLSHS17}. Clearly the ``similar preference'' assumption is too strong. To develop a more sensible recommender system, we make an attempt to explore the social information in a more natural way.


\begin{figure}[t]
\centering
\includegraphics[width=8.0cm]{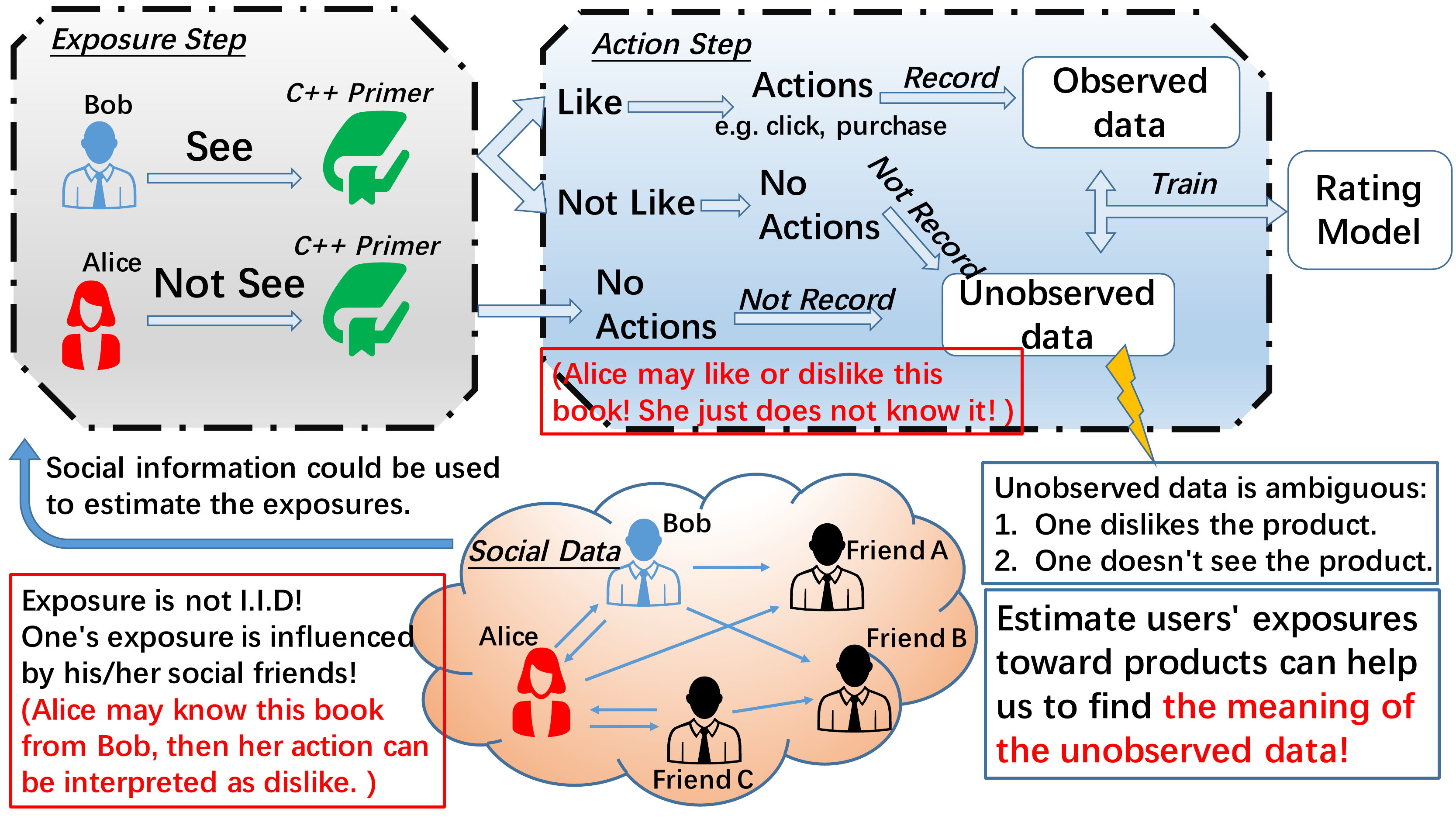}
\caption{An example of user behaviors in a causal perspective. }
\label{model}
\end{figure}

We come to the idea of utilizing social information on the exposure level rather than the preference level. Looking into the process of the user behaviors on websites, we can divide it in two steps: the \emph{Exposure Step} and \emph{Action Step}. In the \emph{Exposure Step} users develop their own exposures on products. Then in the \emph{Action Step} users take actions based on their exposures and preferences towards the products. That is to say, users have to see the products first, then they have the possibility to click or purchase the products based on their preferences. For example, in Figure 1, Bob has seen the \emph{C++ Primer} book, then he takes actions if he likes it or does nothing if he does not like it. Alice does not see this book and does no action to the book. She may like and purchase this book if she has the opportunity to see this book. Note that in the model training, we only have observed data and unobserved data. We do not know which users have seen which products. So the unobserved data is ambiguous with two meanings: 1) one does not like the product; 2) one does not know the product, which makes it hard to learning the user preference. One ideal approach is to model user exposure on each product so we can better capture user preference. Recent studies \cite{Liang2016Modeling,schnabel2016recommendations} mine exposures based on users' inner knowledge and cognitive bias. However, it is nature that user exposures are also influenced by their social friends. In reality social networks often provide plenty of opportunities for users to get information of products from their friends. So we assume that social information influences users on the exposure level, and then influences rating predictions by exposures. This idea gives a more detailed and sensible understanding of social recommendation: the social information is considered on the exposure level and is separated from the rating model. We can apply advanced social network analysis without constraints of rating models.

In this paper, we present a novel social recommendation approach, named ``collaborative filtering with Social Exposure Recommendation'' (SERec), to integrate social exposure into collaborative filtering. Different from traditional social recommendation which fuses the social matrix into the user-item matrix of the rating model, our SERec uses social information to model user exposures on items.
The main contributions of this paper are listed as follows:
\begin{itemize}
\item Our SERec uses social information to estimate user exposures towards items, and then uses them to guide the rating prediction for recommendation. To the best of our knowledge, this is the first work to relax the ``similar preference'' assumption for social recommendation and exploit social information in a modular way.
\item We propose two methods to implement SERec, namely \emph{social regularization} and \emph{social boosting}. \emph{Social regularization} uses matrix factorization to model the social exposure, and utilizes the social information as a regularizer to constrain exposures between users and their friends. \emph{Social boosting} makes use of a general function to compute the social exposure, which can be extend to advanced social network studies.
\item We conduct extensive experiments on four public datasets to verify the efficacy of SERec. The experimental results demonstrate that  SERec outperforms other baseline models. Further study on social exposure shows the effectiveness of social information to user's exposure.
Meanwhile, we also compare the robustness and scalability of \emph{social regularization} and \emph{social boosting}.
\end{itemize}

\section{Related Work}

\textbf{Social Recommendation} utilizes social information to improve the performance of recommender systems. Existing social recommendation algorithms can be roughly divided into two categories: memory-based methods and model-based methods. Memory-based methods first obtain correlated users for a user and then aggregate ratings from the correlated users for the missing ratings \cite{Tang2013Social}. Researchers mainly focus on how to obtain the correlated users (e.g., MoleTrust \cite{DBLP:conf/recsys/MassaA07}, TrustWalker \cite{Jamali2009TrustWalker}). Model-based methods often integrate social information into collaborative filtering for rating prediction. There are mainly two subgroups: ensemble methods \cite{Ma2009Learning,Tang2012mTrust,Guo2015TrustSVD,chaney2015probabilistic} and regularization methods \cite{Jamali2010A,Ma2011Recommender,Tang2013Exploiting,yang2016social}. The common rationale behind these methods is that users' preferences are similar to their friends.
Different from all the above works, our work assumes users are influenced on the exposure level rather than the preference level by their friends due to the information spread by social networks.

\textbf{Recommendation with causal inference} is a newly emerging research area. It aims to alleviate the selection biases of training data for recommendation algorithms caused by the Missing Not At Random (MNAR) problem. Unlike traditional methods \cite{Marlin2009Collaborative,ling2012response,Hern2014Probabilistic} that integrate rating prediction and missing data model into a unified Bayesian model with sophisticated approximate inference, causal inference based methods first compute exposures (or the propensity weights) for each user and then use them to guide rating prediction, which gives more appropriate description of the process as two separate modules. MF-IPS \cite{schnabel2016recommendations} proposes an empirical risk minimization approach to learning the unbiased estimators of user's preferences from biased rating data. ExpoMF \cite{Liang2016Modeling} proposes a Bayesian model to capture propensity score by user exposure. These causal process based methods outperform the state-of-the-art traditional recommendation methods, but they do not take social information into consideration. 
\section{The Proposed Framework}
\subsection{Preliminary and Notation}
First we describe some preliminary of our model and corresponding notation. Let \(U = \{u_{1},u_{2},...,u_{n}\}\) and \(I = \{i_{1},i_{2},...,i_{m}\}\) be the set of users and items respectively. For each user-item pair, we use \(\alpha_{ui}\) to indicate whether user \(u\) has been exposed to item \(i\), and use \(y_{ui}\) to indicate whether or not user \(u\) clicks item \(i\).
Following the definitions of \cite{Liang2016Modeling}, whether a user is exposed to an item follows from a Bernoulli. Conditional on the exposure variable \(\alpha_{ui}\), user's preference follows from a matrix factorization model, which computes \(y_{ui}\) by a multiplication of two latent factors: \(\theta\) and \(\beta\).  The detailed notations are shown below:
\begin{equation}
\begin{split}
&\ \theta_{u} \sim \mathcal{N}(0, \lambda_{\theta}^{-1}I_{k}), \\
&\ \beta_{i} \sim \mathcal{N}(0, \lambda_{\beta}^{-1}I_{k}), \\
&\ \alpha_{ui} \sim Bernoulli(\mu_{ui}), \\
&\  p(y_{ui} = 0| \alpha_{ui} = 0) = 1, \\
&\  y_{ui}| \alpha_{ui} = 1 \sim \mathcal{N}(\theta_{u}^{T}\beta_{i}, \lambda_{y}^{-1}).\\
\end{split}
\end{equation}
Where \(\theta_{u}\) denotes user preference of user \(u\), \(\beta_{i}\) denotes item attributes of item \(i\); \(\mu_{ui}\) is the prior probability of exposure; \(\lambda_{\theta},\lambda_{\beta}\), and \(\lambda_{y}\) are hyperparameters, denoting the inverse variances. \( p(y_{ui} = 0| \alpha_{ui} = 0) = 1\) means that when user \(u\) has not seen the item \(i\), the probability for the user to click the item is zero. As we utilize social information to compute the exposure variable \(\alpha_{ui}\), we call it social exposure in this paper. We list the notation in Table 1.

\begin{table}[t]
\small
\label{table_symbol}
\centering
\begin{tabular}{lccc}
\hline
 Symbol &  Description  \\
\hline
\(y_{ui}\) & the rating user \(u\) gives to item \(i\) \\
\(\theta_{u}\)  & user preference vector of user $u$\\
\(\beta_{i}\) & item attribute vector of item $i$ \\
\(\alpha_{ui}\) & indicates whether user \(u\) has exposed to item \(i\) \\
\(\mu_{ui}\) & prior probability of exposure variable \(\alpha_{ui}\)\\
\(X_{u}\) & user exposure latent vector of user $u$ \\
\(T_{i}\) & item exposure latent vector of item $i$ \\
\(B_{k}\) & trustee specific vector of user $k$ \\
\(\gamma_{i}\) & item exposure bias of item \(i\)\\
\(S\) & social connection matrix  \\
\(\Phi(S)\) & social exposure function  \\
\hline
\end{tabular}
\caption{Notation}
\end{table}

\subsection{SERec Framework}
In the following, we describe our proposed framework, named ``collaborative filtering with Social Exposure'' Recommendation (SERec), which integrates social exposure into collaborative filtering for recommendation. In order to mine different aspects of social exposure, we propose two implementations of SERec and show their graphical representations in Figure 2. SERec has two main components: the \emph{Rating Component} and the \emph{Social Exposure Component}.
The \emph{Rating Component} is a graphical representation of equation (1). It is actually a matrix factorization model for rating prediction: the rating is predicted by taking a inner product of two latent vectors, i.e., \(y_{ui} = \theta_{u}^{T}\beta_{i}\).
The \emph{Social Exposure Component} calculates the exposure priori \(\mu_{ui}\) for each user-item pair. The social information is only used in this component. Unlike traditional social recommendations fusing social information into user latent factors, our SERec model utilizes social information to calculate the exposure variable and leave user latent factors \(\theta_{u}\) ``pure'' -- such factors are derived only from the rating information. This brings two merits: (1) The model is more scalable. In the \emph{Rating Component} \(\theta_{u}^{T}\beta_{i}\) contains essential information for recommendation; If needed, recent matrix factorization methods can be straightforwardly incorporated to improve recommendations, without interfering the \emph{Social Exposure Component}. In the \emph{Social Exposure Component} we can apply sophisticated social models with different assumptions and techniques as the only variable the social information can affect is \(\mu_{ui}\); (2) The modularity of SERec reduces the complexity of objective function, and the parameters are inferred into two steps. The separation of the two components makes parameter updates simple and easy to implement. 

For the \emph{Rating Component}, the log joint probability of exposures \(\alpha_{ui}\) and click \(y_{ui}\) for user \(u\) and item \(i\) is

\begin{equation}
\begin{split}
&\ log \, p(\alpha_{ui}, y_{ui} | \mu_{ui}, \theta_{u},  \beta_{i}, \lambda_{y}^{-1})\\
=&\ Bernoulli(\alpha_{ui} | \mu_{ui}) + \alpha_{ui} log \mathcal{N}( y_{ui} | \theta_{u}^{T}\beta_{i}, \lambda_{y}^{-1})\\
&\ ~~~~~~~~~~~~~~~~~~~~~~~~~~~~~~~~~~+ (1 - \alpha_{ui}) log \amalg[y_{ui} = 0],
\end{split}
\end{equation}
where \(\amalg[b]\) is the indicator function which is \(1\) when \(b\) is true, and \(0\) otherwise. In the setting of a recommender system, we care about the prediction results of \(y_{ui}\) and equation (2) is the main objective function.

\begin{figure}[t]
\centering
\includegraphics[width=8.0cm]{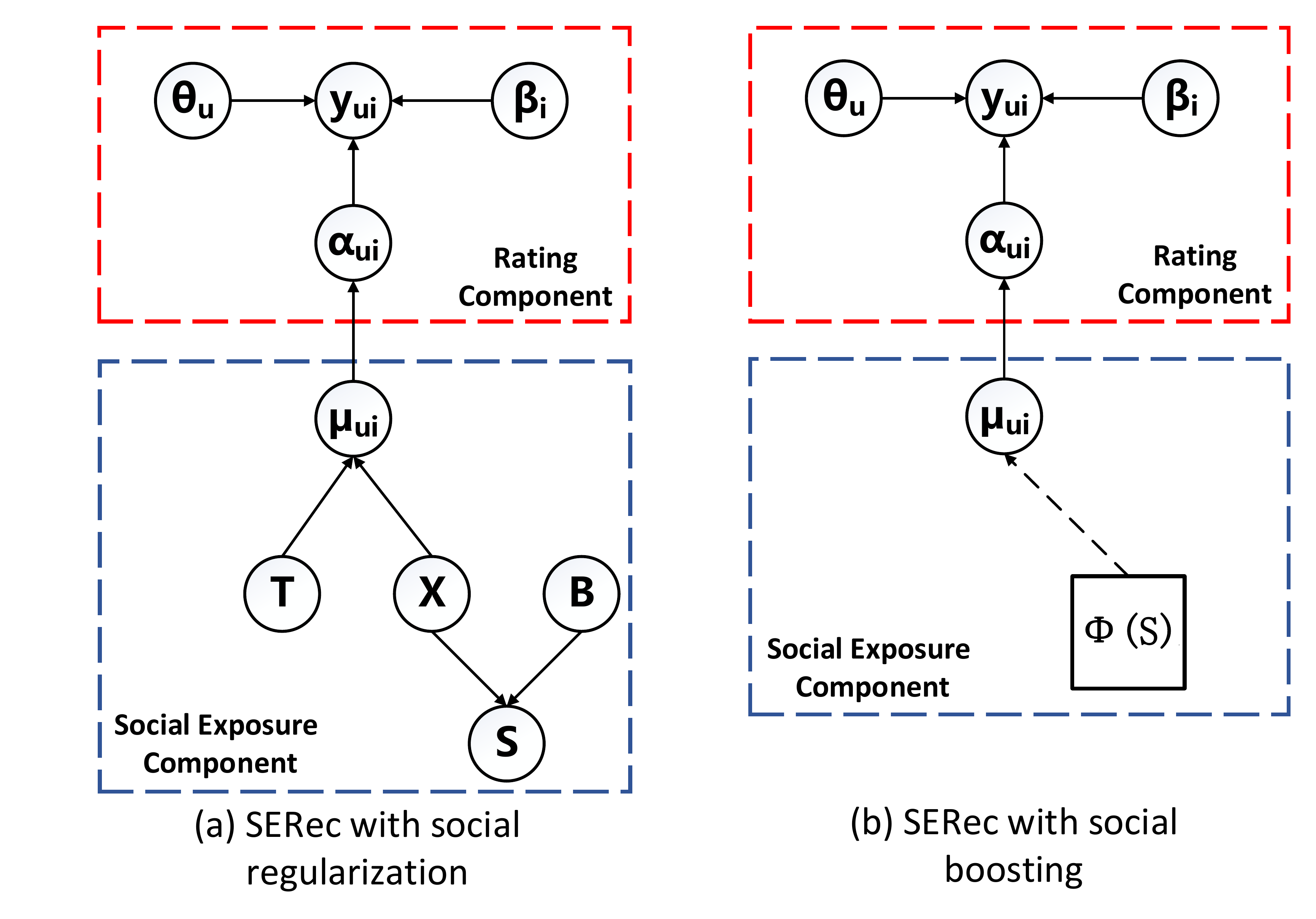}
\caption{Two graphical models of SERec. }
\label{graphic}
\end{figure}

For the \emph{Social Exposure Component}, we propose two methods to deal with the social information, namely \emph{social boosting} and \emph{social regularization}, each with different assumptions. The details of the two methods are shown in the next subsections.

\subsection{Social Regularization}
As regularization methods are effective in traditional social recommendation, a natural idea is to utilize social information as a regularizer in \emph{Social Exposure Component}.  Traditional social recommendation algorithms \cite{Jamali2010A,Ma2011Recommender,yang2016social} assume friends share similarities in rating preferences and introduce social connections as regularization terms in user-item matrix factorizations. Differing from their ideas, we use user connections for social exposure constraints. Here we propose a novel matrix factorization based method to calculate \(\mu\) with social information as regularization. The exposure prior is computed by \(\mu_{ui} = X_{u}^{T}T_{i}+\gamma_{i}\), where \(X_{u}\) is the user exposure latent vector, \(T_{i}\) is the item exposure latent vector, and \(\gamma_{i}\) is the item bias. The social matrix \(S\) is mapped by two latent factors: truster-specific vector \(X\) and trustee-specific vector \(B\). \(X\) and \(B\) characterize the behaviors of ``to trust others" and ``to be trusted by others". We set the \(X\) as a shared variable, so it has two folds of meanings: How a user trusts (or is affected by) others and how the same user has exposure on items. Therefore, we learn the latent factors by minimizing the following objective functions:

\begin{small}
\begin{flalign} \nonumber
 \mathcal{L}_{sr} = & \sum_{ui}(X_{u}^{T}T_{i}+\gamma_{i} - \mu_{ui})^{2} +  \lambda_{sr}\sum_{uk}(X_{u}^{T}B_{k} - S_{uk})^{2}\\
 & ~~+\lambda_{x} \|X_{u}\|^{2}+\lambda_{t} \|T_{i}\|^{2}+\lambda_{b} \|B_{k}\|^{2} + \lambda_{\gamma}\|\gamma_{i}\|^{2}, 
\end{flalign}
\end{small}
where \(\lambda_{sr}\) controls the proportion; \(\lambda_{x}\), \(\lambda_{t}\),  \(\lambda_{b}\), and \(\lambda_{\gamma}\) are regularization parameters to avoid extreme values in the parameters. For those \(ui\) pairs such that \(y_{ui} = 1\), we set \(\mu_{ui} =  \frac{\#users\ who\ rated\ item\ i}{\#total\ users}\) as constraints.
In this way, the estimated exposure component further takes social information into account by constraining \(X_{u}\). For simplicity, we call this method SERec$_{regular}$.

\subsection{Social Boosting}
One potential drawback of SERec$_{regular}$  is that the matrix factorizations on social exposure prior and social matrix assume low rank decomposition structure, which may not well model practical situations.
So we propose a more general method named ``social boosting'', which avoids this potential drawback and can be augmented with recent advanced social models. The intuition of social boosting is simple but reasonable: a consumer's knowledge of products is boosted by his/her friends. S/he gets information about products from friends' discussion and shared feelings. In other words, a consumer will have more exposure on a product if the consumer's friends have interacted with this product. We choose the Beta distribution Beta(\(\alpha_{1}\), \(\alpha_{2}\)) as the item-dependent conjugate prior for \(\mu_{ui}\). The general function of social boosting can be defined as follows:
\begin{equation}
 \mu_{ui} = e_{ui} + \Phi(S),
\end{equation}
where \(e_{ui}\) is the inner exposure of user \(u\) toward item \(i\); \(\Phi(S)\) is the social exposure function that calculates the exposures user \(u\) get from his friends. \(e_{ui}\) can be learned by item popularity, text topics, or user locations \cite{Liang2016Modeling}.
The \(\Phi(S)\) serves as a plug-in function and it is flexible enough to meet different social situations. In this paper we use item popularity to calculate \(e_{ui}\) and set
\begin{equation}
\Phi(S) = \sum_{f\in Friends(u)}s\cdot \mu_{fi},
\end{equation}
where \(s\) is the coefficient of social effects, \(Friends(u)\) denotes the set of the friends of user \(u\). For simplicity, we call this method SERec$_{boost}$.

\subsection{Inference}
We use expectation-maximization (EM) to find the maximum a posteriori of estimates of parameters \(\theta_{u}, \beta_{i}\), and \(\mu_{ui}\). Below we describe the details of each E step and M step.

\textbf{E-step}.
We calculate the expected value of the following log likelihood function.
\begin{equation}
\begin{split}
 L  = &\ \sum_{u,i\in Y^{-}} log (\mu_{ui}\mathcal{N}( 0 | \theta_{u}^{T}\beta_{i}, \lambda_{y}^{-1})+(1-\mu_{ui})) \\
&\ ~~~~~~~~~~~~~~~~~ +\sum_{u,i\in Y^{+}} log (\mu_{ui}\mathcal{N}( 1 | \theta_{u}^{T}\beta_{i}, \lambda_{y}^{-1}))  \\ 
&\ ~~~~~~~~~~~~~~~~~~~~~ -\frac{\lambda_{\theta}}{2}\sum_{u}\theta_{u}^{T}\theta_{u} - \frac{\lambda_{\beta}}{2}\sum_{i}\beta_{i}^{T}\beta_{i}.\\
\end{split}
\end{equation}
where \(Y^{-}\) denotes the set of unobserved user-item pairs, and \(Y^{+}\) denotes the set of observed user-item pairs.

For the exposure variable \(\alpha_{ui}\), we compute the expectation \(E[\alpha_{ui}]\) for every user-item pair. From the definition we can see that \(\alpha_{ui} = 1\) when the click action \(y_{ui}\) is observed. For those unobserved data, we apply the following equation:
\begin{equation}
E[\alpha_{ui}|y_{ui}=0] = \frac{\mu_{ui}\mathcal{N}( 0 | \theta_{u}^{T}\beta_{i}, \lambda_{y}^{-1}))}{\mu_{ui}\mathcal{N}( 0| \theta_{u}^{T}\beta_{i}, \lambda_{y}^{-1}))+(1 - \mu_{ui})}.
\end{equation}
\(\alpha_{ui}\) belongs to a Bernoulli distribution, and \(y_{ui}\) is set to \(0\) when it is not observed. For convenience, we define \(p_{ui} = E[\alpha_{ui}|y_{ui}=0]\). We define \(p_{ui} = 1\) if \(y_{ui} = 1\).

\textbf{M-step}. In this step we aim to find the parameter that maximizes the objective log likelihood function in E-step. We use gradient descent to update the latent collaborative filtering factors \(\theta_{u}\) and \(\beta_{i}\):
\begin{flalign}
\theta_{u} & \leftarrow (\lambda_{y} \sum_{i}p_{ui}\beta_{i}\beta_{i}^{T} + \lambda_{\theta}I_{K})^{-1}(\sum_{i}\lambda_{y}p_{ui}y_{ui}\beta_{i}), \\
\beta_{i} & \leftarrow (\lambda_{y} \sum_{u}p_{ui}\theta_{u}\theta_{u}^{T} + \lambda_{\beta}I_{K})^{-1}(\sum_{u}\lambda_{y}p_{ui}y_{ui}\theta_{u}),
\end{flalign}

\textbf{Update for \(\mu_{ui}\) with Social Regularization}.
Minimizing the objective functions \(\mathcal{L}_{sr}\) needs enormous cost of computation.
To speed up the learning procedure, we use stochastic gradient descent to learn the unknown parameters. At each step we randomly choose a triplet (\(i,u,k\)) and calculate the gradients for update as follows,
\begin{equation}
\begin{split}
&\ \frac{1}{2}\frac{\partial L}{\partial T_{i}} =  (X_{u}^{T}T_{i}+\gamma_{i} - \mu_{ui})X_{u} + \lambda_{t}T_{i},\\
&\  \frac{1}{2}\frac{\partial L}{\partial X_{u}} =  (X_{u}^{T}T_{i}+\gamma_{i} - \mu_{ui})T_{i}  \\
&\ ~~~~~~~~~~~~~~~ + \lambda_{sr} (X_{u}^{T}B_{k} - S_{uk})B_{k}  + \lambda_{x}X_{u},\\
&\ \frac{1}{2}\frac{\partial L}{\partial B_{k}} =  \lambda_{sr} (X_{u}^{T}B_{k} - S_{uk})X_{u} + \lambda_{b}B_{k}.\\
&\ \frac{1}{2}\frac{\partial L}{\partial \gamma_{i}} = (X_{u}^{T}T_{i}+\gamma_{i} - \mu_{ui}) + \lambda_{\gamma}\gamma_{i} .\\
\end{split}
\end{equation}
Then we repeat the process until a minimum is obtained. 
In practice, we find only updating \(\mu\) with social regularization in one EM iteration could give comparable performance and save computation.

\textbf{Update for \(\mu_{ui}\) with Social Boosting}. Maximizing the log likelihood with respect to \(\mu_{ui}\) is equivalent to finding the mode of the complete conditional \(Beta(\alpha_{1} + \sum_{u'}^{U}p_{u'i} + (s -1)\sum_{f}^{ Friends(u)}p_{fi},\alpha_{2} + U -\sum_{u'}^{U}p_{u'i} )\), which is:
\begin{equation}
\mu_{ui} \leftarrow \frac{\alpha_{1} + \sum_{u'}^{U}p_{u'i} + (s -1)\sum_{f}^{ Friends(u)}p_{fi} - 1}{\alpha_{1} +\alpha_{2} + U + (s -1)\sum_{f}^{ Friends(u)}p_{fi}-2},
\end{equation}
where \(U\) is the number of users,  \(s\geq1\) is the coefficient of social effects.

\section{Experiments}
\subsection{Datasets and Settings}

\begin{table*}[t]
\centering
\small
\begin{tabular}{ |l|l|l|l|l|l|l|l|l| }
\hline
\multicolumn{9}{ |c| }{Effectiveness of models }  \\
\hline
Dataset & Metrics & baseMF & RSTE  & TrustMF  & WMF & ExpoMF & SERec$_{regular}$ & SERec$_{boost}$ \\ \hline
\multirow{3}{*}{Lastfm} &  recall@10 & 0.0004 & 0.0045 & 0.0731 & 0.1206 & 0.1483 & 0.2117 & \textbf{0.2159} \\ 
 & recall@50 & 0.0013 & 0.0199 & 0.1898 & 0.3081 & 0.3357  & 0.3990 & \textbf{0.4381}\\ 
 &  MAP@100 & 0.0001 & 0.0018 & 0.0208 & 0.0406 & 0.0366 & 0.0501& \textbf{0.0527}\\
 & NDCG@100 & 0.0008 & 0.0131 & 0.1261 & 0.2385 & 0.2200  & 0.2937 & \textbf{0.3102}\\ \hline
\multirow{3}{*}{Delicious} &   recall@10 & 0.0001 & 0.0005 & 0.0085 & 0.0614 & 0.1818  & 0.1385 & \textbf{0.1934}\\ 
 & recall@50 & 0.0011 & 0.0008 & 0.0160 & 0.2859 & \textbf{0.4448}  & 0.3927 & 0.4442 \\
 &  MAP@100 & 7.3E-5 & 0.0001 & 0.0006 & 0.0331 & 0.0502 & 0.0434 & \textbf{0.0511}\\
 & NDCG@100 & 0.0005 & 0.0005 & 0.0077 & 0.1635 & 0.2501  & 0.2193 & \textbf{0.2542}\\ \hline
 \multirow{3}{*}{Douban} &  recall@10 & 0.0001 & 0.0011 & 0.0810 & 0.1157 & 0.0964 & 0.1132 & \textbf{0.1161}  \\ 
 & recall@50 & 0.0002 & 0.0034 & 0.1878 & 0.2255 & 0.2057  & 0.2302 & \textbf{0.2425}\\
 &  MAP@100 & 5E-6 &  0.0003& 0.0397 & 0.0405 & 0.0403 & 0.0440 & \textbf{0.0489}\\
 & NDCG@100 & 0.0002 & 0.0016 & 0.1265 & 0.1820 & 0.1840& 0.1913  & \textbf{0.2015 }\\ \hline
\multirow{3}{*}{Epinions} &  recall@10 & 0.0006  & 0.0008 & 0.0304 & 0.0385 & 0.0491  & 0.0568 & \textbf{0.0618}\\ 
 & recall@50 & 0.0033& 0.0031 & 0.0801 & 0.1072  & 0.1302  & 0.1347 & \textbf{0.1556} \\
 &  MAP@100 & 0.0002 & 0.0002 & 0.0041 & 0.0071 & 0.0068 & 0.0079 & \textbf{0.0088} \\
 & NDCG@100 & 0.0018 & 0.0016 & 0.0385 & 0.0545 & 0.0634  & 0.0701 & \textbf{0.0777}\\ \hline
 \end{tabular}
\caption{Performance of different models on four datasets. }
\end{table*}

\begin{table}[htb]
\label{table_s}
\centering
\small
\begin{tabular}{|l|c|c|c|c|c|c|c|}
\hline
Dataset&  Epinions  & Delicious & Lastfm  & Douban  \\
\hline
Users (U) & 32,424 & 1,867 &1,892&129,490\\ \hline
Items (V)& 61,274 & 69,223 &17,632 & 58,541 \\ \hline
Ratings (R)& 664,824 & 104,799 &92,834&16,830,839 \\ \hline
R-Density  & 0.03\% & 0.08\% &0.01\%&0.22\% \\ \hline
Social (S)&487,145 &15,328 & 25,434 & 1,692,952\\ \hline
S-Density   & 0.05\% & 0.44\% & 0.71\% & 0.01\%\\ \hline
avg.S & 15.02 & 8.23 & 13.44 &13.07\\ \hline
S-Impact & 3.67\% & 0.67\% & 3.72\% &2.91\%\\
\hline
\end{tabular}
\caption{Data statistics. \(U,V,R,S\) show the counts of each feature; R/S-Density indicates ratings/social links density; avg.S is average social links per user; S-Impact is the average ratio of items with which users' social friends have interacted to the whole item set.}
\end{table}


We consider four public datasets for experiments: \emph{Lastfm}, \emph{Delicious}, \emph{Douban}, and \emph{Epinions}. These datasets are widely experimented by social recommendations. The details of each dataset are listed in Table 3. Note that S-Impact = \(\frac{S}{U}\times\frac{R}{U \times V}\), which is a measurement of social impact by computing the average ratio of items with which users' social friends have interacted to the whole item set. We use Recall, MAP, and NDCG as our metrics. They are common metrics for recommendation and their definitions can be found in \cite{Liang2016Modeling}. To avoid data biases, we randomly select \(70\%\) of each dataset for training, \(20\%\) of each dataset for validation, and the remaining data for testing. To accelerate the experiments, we choose parallel programming and use C++ with OpenMP.

We then choose various prevalent methods for comparison, including: (1) BaseMF \cite{Salakhutdinov2007Probabilistic}, the baseline matrix factorization approach, widely applied as a benchmark; (2) WMF \cite{Hu2008Collaborative}, a standard factorization model for implicit data, which uses a simple heuristic where all unobserved user-item interactions are equally down weighted against the observed interactions; (3) RSTE \cite{Ma2009Learning}, which fuses the users' tastes and their friends' favors in a unified framework; (4) TrustMF \cite{yang2016social}, which utilizes social information as a regularization for recommendation, as one of the state-of-the-art social recommendation models; (5) ExpoMF \cite{Liang2016Modeling}, a new probabilistic approach that directly incorporates user exposure to items into collaborative filtering, which does not utilize social information.

\subsection{Top-N Recommendation}
We use grid search to tune the parameters to achieve the best performance. We set \(K = 10\) for BaseMF, \(K = 10\) and \(\alpha=0.7\) for RSTE, \(K = 30\) and \(\lambda_{sr}=1\) for TrustMF, \(K = 10\) and \(\alpha=0.4\) for WMF, and \(\lambda_{\theta} = 0.01, \lambda_{\beta} = 0.01, \lambda_{y} = 0.01\), and \(K=100\) for ExpoMF. For our methods, we use the above values for \(\lambda_{\theta}, \lambda_{\beta}\), and \(\lambda_{y}\) and set \(K = 20\), \(K_{sr}= 30, \lambda_{x} = 1, \lambda_{t} = 1, \lambda_{b} = 1\), and \(\lambda_{sr} = 5\) for SERec$_{regular}$ and set \(s=5\) for SERec$_{boost}$.

We show the experimental results in Table 2. We can see that SERec$_{regular}$ and SERec$_{boost}$ outperform other methods on all the four datasets, according to all the metrics. PMF has the worst performances because it only uses click data and treats the unobserved and observed data equally. RSTE outperforms PMF slightly since it ensembles social information into matrix factorization with a linear combination. Due to the sparsity problem in both ratings and social links, the performance of RSTE is still rather poor. TrustMF utilizes social information as a regularization and clearly outperforms RSTE, which proves the superiority of regularization methods. Though WMF does not utilize social information, it is superior to TrustMF as WMF treats unobserved user-item interactions with a low confidence. ExpoMF computes exposure latent variable \(\alpha_{ui}\) for every user-item pair, and WMF is a special case of ExpoMF by constraining the  ExpoMF's exposure variable  \(\alpha_{ui}\) to be binary, indicating whether the data is unobserved or observed. ExpoMF is inferior to our models and it does not use social information, which proves the superiority of our SERec models. On the other hand, SERec also outperforms other ways of using social information (RSTE and TrustMF). This verifies that it is essential to find a proper way to exploit social information.  

\begin{figure}[t]
\centering
\includegraphics[width=8.0cm]{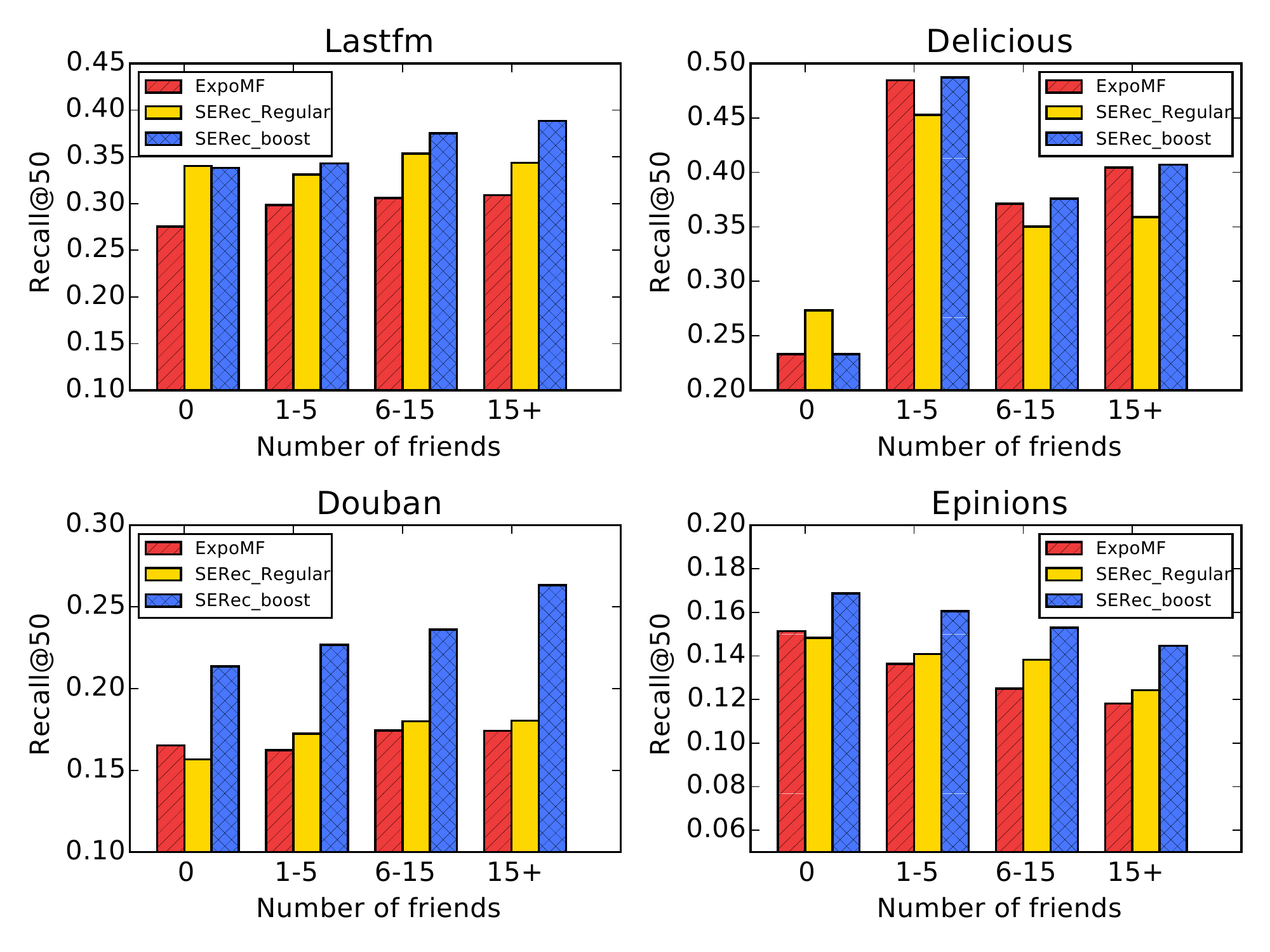}
\caption{Performances of users with different number of friends.}
\label{recall}
\end{figure}

Besides, SERec$_{boost}$ outperforms SERec$_{regular}$ on four datasets. We then compare our two methods  by exploring how well the number of social friends influences the recommendation performance. We divide the users into four groups (namely, \emph{0}, \emph{1-5}, \emph{6-15} and \emph{15+}) based on how many friends they have. We compare the \(recall@50\) among ExpoMF, SERec$_{boost}$, and SERec$_{regular}$ in Figure 3. ExpoMF serves as a baseline because it does not utilize social information. SERec$_{boost}$ outperforms ExpoMF especially for users who have more friends. This conforms to our expectation that users have more exposure when they have more friends (see equation (4)), which increases the possibility of click.
Meanwhile, SERec$_{regular}$ performs more stable across the four groups. Compared to SERec$_{boost}$, we can see that SERec$_{regular}$ is inferior especially on group \emph{6-15} and \emph{15+}. There is an ``exception'' that in dataset \emph{Delicious} SERec$_{regular}$ is worse than ExpoMF except for group \(0\). Note that RSTE and TrustMF also perform very poorly in \emph{Delicious} compared to other datasets. Checking the data distribution of \emph{Delicious} in Table 3, the average links of users is 8.23 and the number of items is 37.07 times that of users, which accounts for the low S-Impact of \(0.67\%\). The S-Impact of \emph{Delicious} is much smaller than that of the other datasets (3.67\%, 3.72\%, 2.91\% respectively), indicating the ``social quality'' of \emph{Delicious} is lower. This is also a sparsity problem: users' friends have few recorded interactions, and accordingly, utilizing them may limit the performance of recommendations. On the other hand, even though the social quality is poor, our methods still outperforms traditional social recommendation methods (RSTE and TrustMF), which proves our ways of utilizing social information. 

\subsection{Analysis of Social Exposure}
In SERec, the rating behavior is controlled by the social exposure. We explore the effectiveness of social exposure variable \(\alpha\) and its prior \(\mu\) with two specific users in dataset Lastfm, namely, User \(A\) and User \(B\). User \(A\) has 14 friends while User \(B\) has no friends. The reason of selection lies in that the avg.links is 13.44 and the social density is 0.71\% (shown in Table 3). So User \(A\) and User \(B\) represent two typical kinds of users: users with average number of friends and users with few or no friends. 

\begin{figure}[t]
\centering
\includegraphics[width=8.0cm]{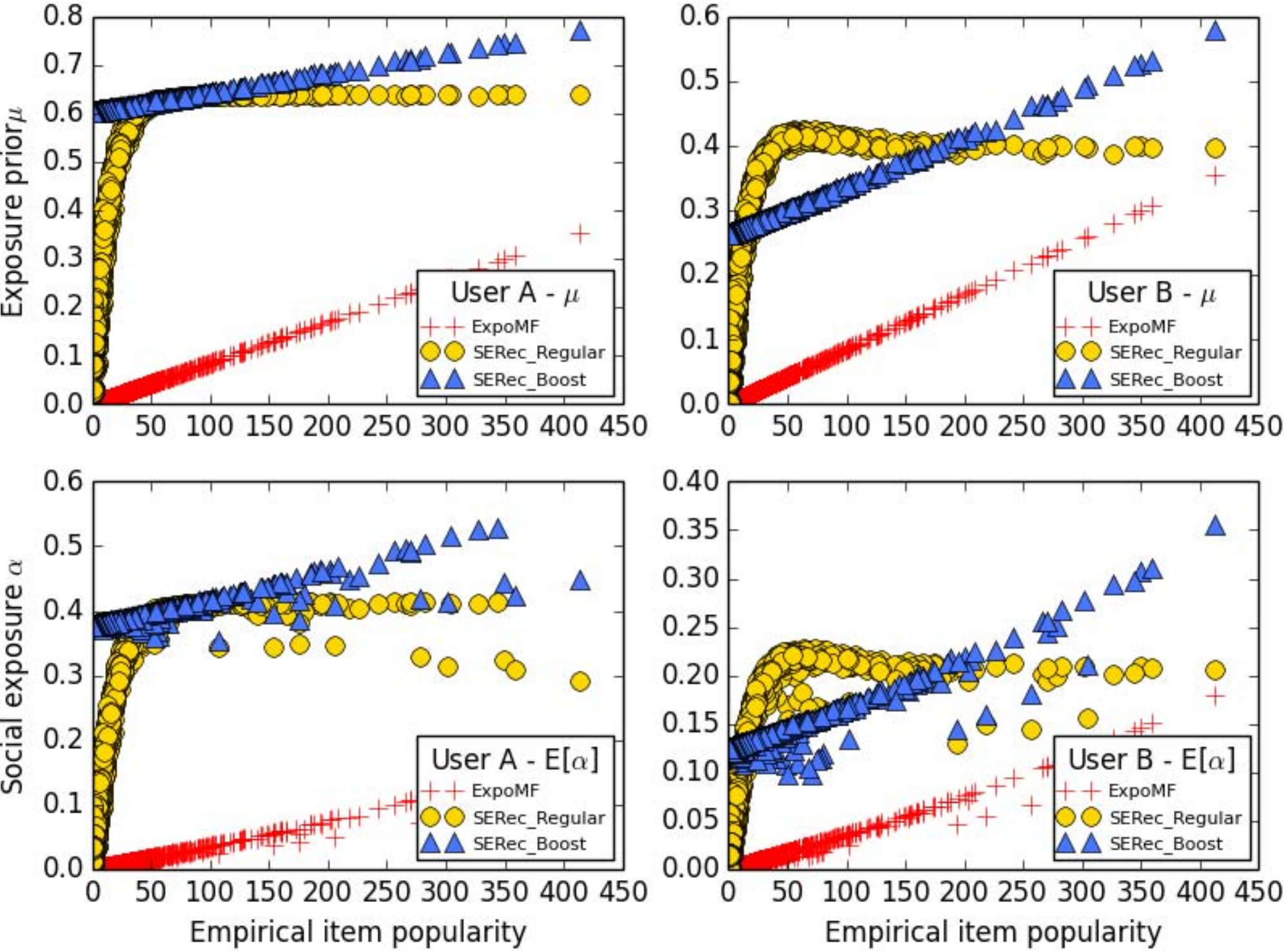}
\caption{Case study of two users: the left column for User A and the right for User B. The Top row shows the exposure prior \(\mu\) and the bottom row shows the expectation of exposure variable \(\alpha\).}
\label{exposure}
\end{figure}

In the Figure 4, we draw the \(\mu\) and  \(E[\alpha_{ui}]\) distributions for the two users. The ExpoMF only utilizes item popularity, so its \(\mu\) curve is a ``perfect'' linear function of item popularity. And this \(\mu\) curve is the same across all users, including User \(A\) and User \(B\). For SERec$_{regular}$, there is a rapid increase of \(\mu\) in Users \(A\) and \(B\) when the item popularity is smaller than 50. After that \(\mu\) becomes stable. We can see the \(\mu\) curve is always above that of ExpoMF, and the \(\mu\) values of User \(A\) is much larger than those of User \(B\). For SERec$_{boost}$, its \(\mu\) curve is always above that of ExpoMF, and the two curves are parallel for User \(B\), because User \(B\) has no friends and the increase of \(\mu\) is caused by the increase of \(p_{ui}\) in equation (10). Meanwhile, the \(\mu\) curve of User \(A\) is less steep than that of User \(B\). This is because: when item popularity is small, \(\sum_{u}^{U}p_{ui}\) is small and the social effect \((s -1)\sum_{f}^{ Friends(u)}p_{fi}\) contributes more to the exposure; when item popularity increases, the \(\sum_{u}^{U}p_{ui}\) increases and the effect of social information is weakened. Accordingly, for the users who have more friends, SERec$_{boost}$ has a larger promotion on their exposures than those who have few friends, and it is more effective for those items with small popularity. Compared to SERec$_{regular}$, SERec$_{boost}$ performs worse for User \(B\) when item popularity varies from 10 to 200. But for User \(A\) SERec$_{boost}$ always performs better. As User \(A\) has more friends than User B, such observations also demonstrate that SERec$_{boost}$ is more effective when users have more friends.


Consider the bottom panels of Figure 4, we can find that the shape of the curve \(E[\alpha_{ui}]\) are similar to that of \(\mu\). As a high \(E[\alpha_{ui}]\) only indicates a high possibility for user \(u\) to see the item \(i\), it does not guarantee that the user will take actions (e.g., click, purchase) on this item. In other words, we can not infer specific recommendation results based only on the social exposure curves. In this section we can conclude that our SERec models improve the user exposure by utilization of social information. SERec$_{regular}$ and SERec$_{boost}$ perform differently due to their different mechanisms, and SERec$_{boost}$ generates higher exposures than SERec$_{regular}$ when users have more friends.

\subsection{Comparison of Robustness and Scalability}
As the \emph{Rating Component} of our two methods is the same, here we focus more on the \emph{Social Exposure Component}.
\subsubsection{Robustness}
We now explore the robustness of SERec to different proportions of social information exploited. We randomly prune some social links between users and use the remaining links in our methods, and observe the effectiveness of recommendation performance. We select two datasets, \emph{Lastfm} and \emph{Delicious}, and show in Table 4 the performance of SERec$_{boost}$ and SERec$_{regular}$ with \(100\%\), \(60\%\), and \(20\%\) of social information utilized. \emph{Lastfm} has a S-Impact of \(3.72\%\), while \emph{Delicious} is of low quality with a S-Impact of \(0.67\%\). We also examine the robustness across different datasets.

\begin{table}[t]
\scriptsize
\begin{tabular}{ |l|l|l|l|l|l| }
\hline
\multicolumn{6}{ |c| }{Performances of SERec$_{regular}$ and SERec$_{boost}$ on the two datasets}  \\
\hline
Lastfm & Metric & P$_{s}$ = 1& P$_{s}$ = 0.6& P$_{s}$ = 0.2& decay ratio\\ \hline
\multirow{3}{*}{Regular.} &Recall@10&0.2110   & 0.2071 & 0.1782& 15.55\%\\ \cline{2-6}
& Recall@50 & 0.3971& 0.3910 &  0.3233 & 18.58\% \\ \cline{2-6}
& MAP@100 & 0.0501& 0.0484 & 0.0417  & 16.77\% \\\cline{2-6}
& NDCG@100 & 0.2937& 0.2824 & 0.2439  & 16.96\% \\\hline
\multirow{3}{*}{Boost.} &Recall@10& 0.2142& 0.1701 & 0.1553  & 27.49\%\\ \cline{2-6}
& Recall@50 &  0.4386 & 0.3933 &0.3801 & 13.34\% \\ \cline{2-6}
& MAP@100 &  0.0527& 0.0501 & 0.0407 & 22.77\% \\\cline{2-6}
& NDCG@100 & 0.3102& 0.2958 & 0.2225  & 28.27\% \\\hline \hline

Delicious& Metric & P$_{s}$ = 1& P$_{s}$ = 0.6& P$_{s}$ = 0.2& decay ratio\\ \hline
\multirow{3}{*}{Regular.} &Recall@10&  0.1385& 0.0802 & 0.0701 & 49.39\%\\ \cline{2-6}
& Recall@50 & 0.3927 & 0.2398 & 0.2020 & 48.56\% \\ \cline{2-6}
& MAP@100 &  0.0434 & 0.0344 & 0.0286 & 34.10\% \\\cline{2-6}
& NDCG@100 &  0.2193& 0.1207 & 0.1007 & 54.08\% \\\hline
\multirow{3}{*}{Boost.} &Recall@10&  0.1934 & 0.1693 &0.1426 & 26.27\%\\ \cline{2-6}
& Recall@50 & 0.4442& 0.3866 & 0.3739  & 15.83\% \\ \cline{2-6}
& MAP@100 & 0.0511& 0.0429 & 0.0399  & 21.92\% \\\cline{2-6}
& NDCG@100 &  0.2542& 0.2406 &0.2112  & 16.92\% \\\hline
\end{tabular}
\caption{Analysis of robustness. P$_{s}$ is the proportion of used social information. Decay ratio means the decay from P$_{s}$ = 1 to P$_{s}$ = 0.2. }
\end{table}

In Table 4 we can see that the performance of SERec$_{boost}$ has almost a linear positive correlation with the percentage of utilized social information both in \emph{Lastfm} and  \emph{Delicious}. And the decay ratio is about \(20\%\) in both datasets. SERec$_{boost}$ is an incremental method that utilizes the linear sum of one's social connections. It reduces to ExpoMF if no social information is available. But in SERec$_{regular}$ the performance varies across datasets. It has a low decay ratio around \(17\%\) in \emph{Lastfm} and a high decay ratio around \(50\%\) in \emph{Delicious}, due to the different S-Impact of two datasets. When the amount of utilized social information decreases, the effectiveness of regularization is weakened and the model may suffer the overfitting issue, accounting for the decrease in the performance of SERec$_{regular}$.
To conclude, it is flexible for SERec$_{boost}$ to deal with different amount of social information, and its performance is more stable across different datasets. SERec$_{boost}$ is more robust than SERec$_{regular}$ to how much social information is used.

\subsubsection{Scalability}We analyze the scalability of SERec$_{boost}$ and SERec$_{regular}$ in three aspects. (1) \textbf{Time cost}: The time complexity of the \emph{Rating Component} in SERec$_{regular}$  and SERec$_{boost}$ is the same. The cost is \(O(UVt)\), where \(U\) is the number of users, \(V\) is the number of items, and \(t\) is the required iterations.
For the \emph{Social Exposure Component}, the complexity of SERec$_{regular}$ is \(O(td(R +S))\), where \(d\) is the dimensionality of feature vectors, and \(R\) and \(S\) are the numbers of observed ratings and observed social links, respectively. The complexity of SERec$_{boost}$ is \(O(UVf)\), where \(f\) is the average number of friends. Because of the severe sparsity problem shown in Table 2, we can see that \((R +S)<<UV\). Besides, SERec$_{regular}$ uses stochastic gradient decent for inference. So we can conclude that SERec$_{regular}$ is more computationally efficient. Meanwhile, the above analysis assumes that our methods run in a sequential fashion, which makes \(O(UV)\) a big overhead. But due to the independence of users in the EM procedure, our methods can be easily devised in a parallel form. In experiments we achieved a 30\(\times\) speed up with C++ and OpenMP. Table 5 shows the elapsed time for training SERec$_{regular}$ and SERec$_{boost}$, and SERec$_{regular}$ is indeed faster. (2) \textbf{Space cost}:  SERec$_{regular}$ calculates \(\mu_{ui}\) with a matrix factorization method, and the memory overhead is \(O(Ud+ Vd)\), where \(d\) (\(d<<U,V\)) is the dimension of feature vectors. But SERec$_{boost}$ has to store \(\mu_{ui}\) for every possible user-item pair. Accordingly its space cost is \(O(UV)\), which will lead to a big memory overhead when there are millions of users and items. How to reduce its space cost is worth further exploring. Compared to SERec$_{boost}$, the cost of SERec$_{regular}$ is relatively small.
(3) \textbf{Model extension}: Potential extensions of SERec$_{regular}$ may make use of alternative methods for matrix factorization. SERec$_{boost}$ does not rely on matrix factorization and can handle more sophisticated situations given that \(\Phi(S)\) is properly defined. For example, in this paper we assume that all friends are equally close; we can find a way to measure the closeness among friends as closer friends tend to have more influence regarding item exposure. Social information based exposure may be improved by further resorting to recent social network analysis techniques (e.g., social contagion \cite{DBLP:conf/aaai/YangJWT16} and social structural influence \cite{DBLP:conf/aaai/ZhangTZMLSHS17}).
To summarize, SERec$_{boost}$ costs more space and time, but it performs better and allows straightforward extensions. SERec$_{regular}$ is more competitive in space and time cost.

\begin{table}[tph]
\centering
\scriptsize
\begin{tabular}{ |c|c|c|c|c|}
\hline
\multicolumn{5}{ |c| }{Runtime Comparison (In Seconds)  }  \\
\hline
Method  & Lastfm& Delicious& Douban & Epinions\\ \hline
Regular.  & 66 \(\pm\) 0.66 & 234 \(\pm\) 3.66  & 122,206 \(\pm\) 33.0 & 2,339 \(\pm\) 13.5\\ \hline
Boost.  & 124 \(\pm\) 2.5& 549 \(\pm\) 5.0  & 330,431\(\pm\) 45.0 & 4,738 \(\pm\) 19.0 \\
\hline
\end{tabular}
\caption{Runtime comparison of two methods of SERec. }
\end{table}

\section{Conclusion and Future Work}
In this paper we proposed a novel approach to social recommendation, called SERec. It integrates social exposure into collaborative filtering for recommendation in a sensible, modular way. We utilized social information in two ways, namely, social regularization and social boosting. Empirical results on four benchmark real-world datasets show that our methods clearly outperform alternatives. Further experimental results demonstrate the influence of social information on user exposures and the usefulness of such information for better recommendation. We also compared the robustness and scalability of the two proposed implementations of SERec. In the future, we will investigate how to leverage alternative social analysis models to better capture users' exposures. Another future line of research is to discover and make use of other sources of information that helps better model the process of users' decision making.

\section{Acknowledgments}

This work was supported in part by the National Natural Science Foundation of China (NO. 61379034 and U1509221), the National Key Technology R \& D Program (NO. 2014\-BAH28F05, 2015BAH07F01), the Zhejiang Province Science and Technology Program (NO. 2017C03044), and the Guangdong Province Science and Technology Program (NO. 2014B040401005). KZ would like to acknowledge the support from NIH-1R01EB022858-01, FAIN-R01EB022858, NIH-1R01LM012087, NIH-5U54HG008540-02 and FAINU54HG008540. The content is solely the responsibility of the authors and does not necessarily represent the official views of the National Institutes of Health.

\bibliography{aaai18_sec}
\bibliographystyle{aaai}

\end{document}